# First principle based phase stability in PMN-*x*PT

M. Sepliarsky[1] and R. E. Cohen[2]

(1)   Instituto de Física Rosario, Universidad Nacional de Rosario – CONICET. 27 de Febrero 210 bis, Rosario, Argentina.  e-mail: sepli@ifir-conicet.gov.ar

(2) Geophysical Laboratory, Carnegie Institution of Washington, 5251 Broad Branch Road, NW, Washington DC 20015, U. S. A

We have applied a shell model potential developed by fitting first principle results to describe the behavior of the relaxor-ferroelectric PMN-*x*PT as function of concentration. The solid solution exhibits three regions with different characteristics according to Ti content in agreement with the experiments. At low Ti the polar and structural disorder dominates, and the symmetry is rhombohedral. The intermediate region is ferroelectric with easy rotation of the polarization direction, and the symmetry appears orthorhombic. Finally, in the high Ti content region, the solid solution adopts a ferroelectric behavior similar to PT, with tetragonal symmetry.

# I. Introduction

Relaxor ferroelectric solid solutions like $(1-x)$PbMg$_{1/3}$Nb$_{2/3}$O$_3$-$x$PbTiO$_3$ (PMN-$x$PT) or $(1-x)$PbZn$_{1/3}$Nb$_{2/3}$O$_3$-$x$PbTiO$_{1/3}$ (PZN-PT) are technological important compounds because they posses electromechanical properties superior to the conventional PZT [1]. As a consequence, the basic understanding of the origin of their properties has becoming an intensive area of research [2-10]. Much progress has been made from theoretical and experimental studies, but other many aspects of such rich behavior still not clearly understood due to the complexity of these materials. The behavior depends on the interplay of many effects like composition or chemical and structural disorder, and involves phenomena in a wide range of lengths and timescales.

PMN-$x$PT is a solid solution of the relaxor PMN and the ferroelectric PT. It has a perovskite type of structure ABO$_3$ with lead at the A site while equivalent B-sites can be occupied by Mg$^{+2}$, Nb$^{+5}$ or Ti$^{+4}$. The solid solution exhibits a variety of physical properties as function of the concentration, and the phase diagram can be described in regions according to the PT content. Similar to other relaxors, PMN is a disordered crystal with a broad-frequency dependant peak in the dielectric constant. It is generally accepted that relaxor behavior is associated with formation, growing, and freezing of disordered polar regions of nanometric size or polar nanoregions (PNR) (see Ref. [4] for a review) while PNR are related with a degree of short-range chemical order [9]. The addition of PT in the composition produces a transition to a ferroelectric phase although relaxor characteristics persist in this ferroelectric zone with rhombohedral symmetry. The origin of the crossover between these two states is uncertain, and it could be produced either by the growing of PNR or by interactions between them. This low Ti region remains up to $x$=0.3. For concentrations above $x$=0.37 the solid solutions is tetragonal with a behavior similar to PT. The region in between is a transition zone denominated morphotropic phase boundary (MPB) which is of a particular interest since it is near this region where the largest electromechanical response takes place. In this region the polarization can easy rotate between the rhombohedral and tetragonal structure mediated by lower symmetry monoclinic or orthorhombic phases. The interplay of an intrinsic behavior with chemical order and the microstructure adds more complexity to the behavior in the MPB zone.

The description of the behavior of complex solid solutions like PMN-$x$PT requires elucidating process at different lenghscales, and it is necessary to link result from first principle calculations, atomic level models and mesoscopic models in a consistent multiscale description. In this way, accurate first-principles results can be transfer to simple models, like effective Hamiltonian or potential models, to investigate properties as function of composition, temperature, electric field and/or pressure. In present work we apply a classical model potential obtained from first principle results to study the evolution of PMN-$x$PT as function of the concentration. Our simulations show that the model is able to describe the behavior of the solid solution in whole range of concentrations in good agreement with the experimental evidence, and it allows relating the macroscopic behavior with the evolution of the local structure.

## II. Model and computational details

Atomic level simulations with the shell model offer a reliable description of ferroelectrics materials, and it has been widely used in the study of pure compounds and solid solutions [11]. In order to takes into account the atomic polarizability, the approach describes each atom as two charged and bounded particles: a core and a shell. The model also includes electrostatic interactions among cores and shells of different atoms, and short range interactions between shells. In this work we use a shell model where core and shell are linked through an anharmonic spring, $V(w)=1/2k_2w^2+1/24k_4w^4$, where $w$ is the core-shell displacement, and Rydberg potentials, $V(r)=(A+Br)e^{-r/\rho}$, are used for the short range interactions between the A-O, B-O and O-O pairs. The input data to adjust the parameters corresponds to first principle results within the local density approximation (LDA). The model was developed by fitting both end members PMN and PT [12], under the assumption of the transferability of interatomic potentials. That is, the model description considers that interactions present in both compounds are equal. In this way, it is possible to study the solid solution in the whole range of concentrations without having to add any extra parameter dependent on concentration or atomic ordering. The model, which parameters are shown in Table 1, was applied to predict elastics constant in PMN [13].

We used the potential to determined relaxed structures of PMN-$x$PT as a function of composition through zero-temperature MD simulations. We used the program DL-POLY [14] within the constant (N, σ, T) ensemble at intervals of 10% in concentration. The runs were performed in system sizes of 12x12x12 unit cells (8640 atoms) with periodic boundary conditions. A fraction of the atomic mass is assigned to the shell to permit a dynamical description of the adiabatic condition [15]. A mass of 7 a.u. is assigned to the Pb, Ti, and Nb shells and 2 a.u. to the O and Mg shells. The time step is of 0.4 fs, and each run corresponds to 50000 steps. The final structures are determined by successive heats and quenches in order to avoid high energy metastable states. Final forces on individual ions are lower than 0.01 eV/Å. Nevertheless, our relaxed configurations are not necessary the ground state of the system.

One central point in the simulation is the treatment of the order at the perovskite B-site, which can occupy by $Nb^{+5}$, $Mg^{+2}$ or $Ti^{+4}$. The current model that fits the existing experimental data is the existence of chemical ordered regions (COR) immerse in a disordered matrix [9]. The COR in PMN have a distribution according to "random site model" (RSM) [16]. In the RSM B-cations display a rock salt type order, β'β", where one of the sublattice is occupied completely by Nb and the other contains a random distribution of Mg and the remaining Nb.

We do not include such nanoregions in our simulations, but rather we occupied the B-sites according to the random site model for Mg and Nb. In addition Ti atoms are randomly distributed, preserving the neutrality of the system. We therefore do not expect to see relaxor behavior, but rather the properties of short-range ordered PMN-$x$PT. We performed simulations for three different arrangements of ions in the supercell (termed 'layouts' below), with different random assignment of positions to analyze their influences on the properties.

## III. Results

### Macroscopic behavior

Figure 1 shows the volume dependence with the composition in PMN-$x$PT at T=0K obtained with the model. Differences between layouts at each concentration are lower than 0.2%, suggesting that in this case results do not depend on the particular B-cation assignment. The figure also includes experimental data at different concentrations from the literature [8,10,17-20], and LDA values for PMN and PT. The volume behavior with composition is in qualitative agreement with experiments. In the model, the volume reduces from 64.10 Å in PMN to 61.36 Å in PT whereas experimentally the change is from 66.3 Å [17] to 62.6 Å [20] respectively. The model results are consistent with LDA, in which the volume varies between 63.9 Å and 64.4 Å depending on the symmetry of the structure [21,22], while the volume for PT is 60.37 Å [23]. The underestimation of the experimental volume with the interatomic potential can be attributed to the difficulty LDA has predicting equilibrium volume in ferroelectrics. In the case of PT, the model gives a larger value than LDA. Thus, fortuitously making the volume of the solid solution closer to experiments for higher concentrations.

Figure 2 shows the evolution of lattice parameters (a) and components of polarization (b) as function of the concentration. For pure PMN limit, the MD simulations are macroscopically cubic and nonpolar for the three layouts. Our results are in agreement experimental observations [5], although we could expect a polar state in supercells ordered according to the random site model. When a small amount of Ti is added ($x$=0.1), a relative small polarization shows up roughly along [111] direction while the cubic structure is slightly distorted. In this low Ti content region, the solid solution has rhombohedral structure while the magnitude of the polarization P is growing with $x$. The layouts have a similar behavior here. At $x$=0.3 the solid solution develops a clear ferroelectric state, and it begins a transition zone between the regions with rhombohedral and tetragonal symmetry. The structure in the region is not clearly defined, and differences in lattice parameters and polarization components between layouts are significative. The region extends up to $x$=0.5, and it corresponds to the MPB. The simulated MPB is broader than the slim region seen experimentally [24], although we can expect the region shrinks in simulations at finite temperature. Finally, for concentration above $x$=0.6 the structure becomes tetragonal with a polarization along [001] direction. In this region, both tetragonal distortion and P grow as Ti content increase, reaching their maximum values at the ending composition. In this region, the behavior does not depend on the layouts.

Our simulations show a compositional induced MPB which is consistent with the polarization rotation mechanism. The polarization magnitude shows only a small change with composition. It changes from approximately 44 $\mu C/cm^2$ at $x$=0.3 to 48 $\mu C/cm^2$ at $x$=0.5. The polarization direction and cell deformation are compatible (Table 2), We also observe that layouts at the same concentration are very close in energy, although they do not have necessary the same symmetry. Since layouts have a similar type of order and they display a similar behavior in the low and high Ti regions, our results reflect a delicate stability between difference phases where small perturbations can change the relative stability between them. In our simulations the perturbations are related to small difference in the random assignment of B-cation or to the relaxation process. Therefore, we conclude that the behavior of the simulated MPB region are due the existence of an underlying flat energy surface as it is require for polarization rotation effect [25]. We estimate the probable polarization path across the MPB from the various phases obtained in present simulations. The phases are rhombohedral (R), monoclinic type $M_B$, orthorhombic (O), monoclinic ($M_C$), tetragonal (T), and triclinic (Tr). We do not find any evidence in present study suggesting the presence of a $M_A$ phase as in the case of PZT. Instead, we find an O phase present at the different concentrations and layouts. The O phase links the R and T phase between two paths. The first corresponds to a R-$M_B$-O sequence where the polarization can rotate in a (0-10) plane, and the second corresponds to a O-$M_C$-T sequence with a polar rotation in a (001) plane. Note that the low symmetry Tr phase is away from the probable path, and the presence may be due to numerical uncertain or size effects. Therefore the estimated zero temperature polarization path across the concentration-induced MPB corresponds to a R-$M_B$-O-$M_C$-T phase sequence in coincides with experimental observations in PMN-$x$PT [26] and PZN-$x$PT [27].

**Microscopic behavior**

The macroscopic behavior of the lattice parameters and polarization can be connected with the behavior at microscopic level. For this purpose we choose the pair distribution function (PDF) and the local polarization distribution function ($W(p)$). The PDF gives information on the local structure, and it allows comparing results of the model with experimental data. Beside, $W(p)$ allows to link our results with models where the polarization is the order parameter. Here, we define the local polarization as the polarization of a perovskite cell centered at B-site

Although pure PMN exhibits a cubic cell, the PDF obtained from the simulations (Figure 3) indicates that ions are displaced in different directions from the high-symmetry perovskite positions. Since thermal effects are not present, the width of the peaks reflects the range of distances the pairs can have and the randomness in atomic displacements from the ideal sites. Model results are in a general good agreement with EXAFS analysis [7] and LDA calculations [28], although bond distances in present simulations are slightly shorter due to the smaller volume. From the different pair contribution to the PDF we distinguish that the peak at approximately 2 Å corresponds to the superposition of Nb-O and Mg-O bonds, and the apparent splitting of ≈ 0.1 Å (shoulder at right side) just reflects a larger repulsion of the Mg-O pair respect to the Nb-O one. The peak at 2.83 Å correspond to Pb-O and O-O pairs, while shoulders at both sides are consequence of the amplitude (splitting) of distances that Pb-O bond takes. Large Pb off-centering is common in relaxors and ferroelectric perovskites [7,29,30], and it provides a strong contribution to local polarization. From this bond, we estimate an average Pb displacement with respect of the center of the $O_{12}$ cage of 0.29 Å and a maximum displacement of 0.5 Å, which is a little underestimated with respect to experimental average value of 0.4 Å [30]. The peak at 3.46 Å corresponds to Pb-Mg/Nb pair, where the bond distance of Pb-Nb pair is slightly larger than the Pb-Mg one due to the more repulsive character in agreement with DFT results [28]. Finally, the peaks at 4 Å and 5.65 Å indicate that the overall cubical symmetry is maintained despite the variety of atomic displacements. The good PDF description obtained in our simulations indicates that the model describes reasonably well the interplay of the different interactions which drives the resulting equilibrium structures [28].

Local polarizations are distinguished from their macroscopic averages. Whereas PMN is non-polar, individual cells are polarized with an average value of 25 $\mu C/cm^2$ and a maximum of 40 $\mu C/cm^2$. Pure PMN therefore behaves like a polar glass. The presence of local polarizations in PMN is seen also in LDA computations for 60-atom supercell, which showed an average local polarization of 38 $\mu C/cm^2$, and a local maximum of 67 $\mu C/cm^2$ [28]. The randomness in atomic displacements produces polarizations in random directions, and *W(p)* along the different directions exhibit a similar broad distribution centered at *p*=0 as it is shown in Figure 4. There is no evidence of the presence of PNR in any of the simulated PMN configurations, neither in PDF nor in *W(p)*, but this is not surprising given the theory that PNR form from local ordered regions, which we did not include in our simulations [31]. In the low Ti region it is hard to distinguish changes in the PDF with respect to PMN. However we can obtain a better picture of local effects from the contributions of individual pairs to the local structure as it is observed in the inset of Figure 5. According to the radial distribution function for Nb-O, Mg-O and Ti-O pairs, each bond has a different distance, but while the first two display a unimodal distribution, the Ti-O pair splits. The ability of Ti atoms to move off the center of oxygen octahedral and form short and large bounds in perovskites results from the balance between electrostatic and short range interactions and it favors the ferroelectricity [23].

The incipient macroscopic polarization observed is reflected as a peak in $W(p)$ (Figure 6), although the broad and asymmetric distribution indicates that polar disorder still dominating the behavior of the system. Interestingly, the net polarization does not come just from Ti-cell, and the three type of B-cell contributes in a rather similar way (inset Figure 6). A macroscopic polarization involves a collective behavior. In a normal ferroelectric, for instance, a stable polarization requires chains of correlated atomic displacements of approximately 5 unit cells (20 Å) [32]. Our simulations suggest that the polar state in PMN-$x$PT originates in a similar type of cooperative behavior where the local order around Ti propagates in the structure favoring the polar reorientation of neighboring cells. The development of polar order in PMN-$x$PT by Ti substitution is in good agreement with experiments. High resolution x-ray diffraction studies show the development of a ferroelectric state in PMN-$x$PT for concentration as low as 0.05 [10]. In this case the relaxor to ferroelectric state was explained in terms of the increment of the size of PNR. Since PNR are not present in the simulations, our results rather show a transition from a disorder polar state to an ordered one triggered by Ti concentration.

Local order is stronger in MPB region, and there are better defined bond distances. However the position of the peaks depends on the particular phase. The PDF for PMN-0.3PT (Figure 7) clearly shows the splitting of the peak corresponding to the Nb/Mg/Ti-O bond, where each pair splits and not only the Ti-O one. Besides, B-cation influence on the oxygen cage can be distinguished from the shift of the position of the peaks according the pair corresponds to Ti, Nb or Mg respectively. Signs of the ordering in the structure are also observed in the splitting of peaks related with Pb: Pb-O and Pb-Nb/Mg/Ti. The higher degree of order is also observed in *W(p)*, where the components of the local polarizations are uniformly distributed around mean values (along [111] direction in the case of Figure 8). The width of the distribution is an indication of the degree of disorder that is present in the system.

In the high Ti region, both PDF and *W(p)* display sharper peaks, but structural and polar disorder remains in the solid solution until Mg and Nb are completely substituted by Ti. The PDF for *x*=0.9 (Figure 9) shows four different peaks for B-O pairs where each one corresponds mainly to Ti-O, Nb/Mg-O, Ti/Nb-O and Mg-O respectively. In addition, peaks at 3.86Å and 4.18 Å correspond to the tetragonal lattice parameters. Local polarizations are more concentrated around to their average values as it is shown in *W(p)* in Figure 10, and there are also small peaks along both polar and non-polar directions. These last peaks reflect the presence of Nb and Mg in the composition, and their positions are related with the effects produced by isolated Nb and Mg acting as impurities in pure PT. When a Nb atom substitutes a Ti in PT, the bigger charge produces a larger displacement along polar direction from the center of oxygen octahedral. In addition, the stronger Pb-Nb repulsion increases the cell volume and reduces c/a. As result, the polarization of the cell reduces to 57.3 $\mu C/cm^2$ while neighboring cells are little affected. Whereas the upper and lower cells diminish slightly their values, those that are to the sides display a small component ($2\mu C/cm^2$) against the impurity. On the other side, a Mg in a Ti matrix experiences a smaller of center displacement along polar axis due to its more ionic character, and its cell polarization slightly decreases to 63.7 $\mu C/cm^2$. The changes in volume and in c/a are similar to the ones produced by the Nb although less pronounced. Nevertheless, the stronger Mg-O repulsion expands oxygen octahedral and produces significative changes in neighboring cells. The polarization of the upper cell is reduced to 52.4 μC/cm², while lateral cells present a component normal to the polar axis and pointing towards the Mg of 6.5 μC/cm².

## IV.  Conclusions

In this work we showed that a classical atomic level model fitted to first principle results successfully describes the complex PMN-$x$PT as function of the composition. The behavior of the solid solution results from the competition between the end members. While PMN favor a disordered polar state with an average cubic and non polar structure, PT favors an ordered tetragonal structure. As consequence, the solid solution behaves as a complex system where both order and disorder characteristics coexist. In the low Ti region, disorder dominates although a weak ferroelectric order is present. The degree of order in MPB region is enough to develop a ferroelectric phase but the level of disorder does not allow polar alignment a along a preferential direction. Finally, the order overcomes disorder, and a PT-likes behavior is present in the high Ti region.

Many factors converge to successfully model description including: the combination of first principle calculation with a classical model to perform simulations with a large number of atoms, the intrinsic ability of the shell model to describe ferroelectrics, and the validity of the assumption of transferability of interatomic potential even in this case where heterovalent substitutions are involved. The ability to describe zero temperature properties of the solid solution makes the model suitable to investigate many problems related to the behavior of the solid solution at different concentrations.

**Acknowledgments**

This work was sponsored by Consejo Nacional de Investigaciones Científicas y Tecnológicas de la Republica Argentina (CONICET) and the US Office of Naval Research grant number N00014-07-1-0451.

**Table 1**: Shell model parameters for PMN and PT. Units of energy, lengh and charge are given in eV, Å, and electrons.

| Atom | Core charge | Shell charge | $k_2$ | $k_4$ |
|---|---|---|---|---|
| Pb | 5.1464 | -3.3506 | 75.35 | 26896.6 |
| Mg | 2.4628 | -0.1046 | 85.75 | 0.0 |
| Nb | 5.5279 | -2.3798 | 713.64 | 3983.9 |
| Ti | 9.7297 | -6.8449 | 1937.88 | 0.0 |
| O | 0.7057 | -2.2659 | 26.48 | 1324.4 |
| Short range | A | B | $\rho$ | |
| Pb-O | 6291.34 | 296.28 | 0.265259 | |
| Mg-O | 1042.14 | 63.48 | 0.315781 | |
| Nb-O | 1508.13 | 3.78 | 0.298781 | |
| Ti-O | 1416.39 | 3.68 | 0.290791 | |
| O-O | 283.41 | -103.27 | 0.520557 | |

**Table 2**: Structural characteristics obtained with the model for PMN-*x*PT in the MPB region. Note that the symmetries of the phases are approximate.

| Layout | Phase | ΔE (meV/atom) | $P_x$ | $P_y$ | $P_z$ |
|---|---|---|---|---|---|
| 30% | | | | | |
| 1 | R | 0.00 | 22 | 23 | 24 |
| 2 | $M_B$ | +0.50 | 18 | 24 | 24 |
| 3 | O | +2.43 | 1 | 28 | 28 |
| 40% | | | | | |
| 1 | Tr | 0.00 | 20.2 | 24.5 | 29.6 |
| 2 | O | -0.04 | 0.1 | 30.6 | 30.4 |
| 3 | O | +0.70 | 1.7 | 30.2 | 30.3 |
| 50% | | | | | |
| 1 | O | 0.00 | 3.3 | 30.9 | 33.0 |
| 2 | $M_C$ | +0.02 | 1.3 | 28.0 | 35.6 |
| 3 | T | +0.36 | 0.7 | 2.8 | 42.9 |

**Figure captions**

Figure 1: Volume as function of Ti content in PMN-$x$PT obtained from the simulations at T=0K. The figure includes experimental values from the literature [], and LDA results from PMN and PT.

Figure 2: Lattice parameters (a), and polarization (b) as function of Ti-content in PMN-xPT at T=0 K obtained from model simulations. The error bars represent difference between layouts.

Figure 3: Pair distribution function for PMN at 0 K obtained from the simulations.

Figure 4: Polar distribution function for PMN at 0 K obtained from the simulations.

Figure 5: Pair distribution function for PMN-0.2PT obtained from the simulations. The inset shows the radial distribution function for the nearest B-O pairs.

Figure 6: Polar distribution function for PMN-0.2PT obtained from the simulations. The inset shows *W(p)* along the z-component for Nb, Mg and Ti cells.

Figure 7: Pair distribution function for PMN-0.3PT obtained from the simulations.

Figure 8: Polar distribution function for PMN-0.3PT obtained from the simulations.

Figure 9: Pair distribution function for PMN-0.9PT obtained from the simulations.

Figure 10: Polar distribution function for PMN-0.9PT obtained from the simulations.

Figure 1

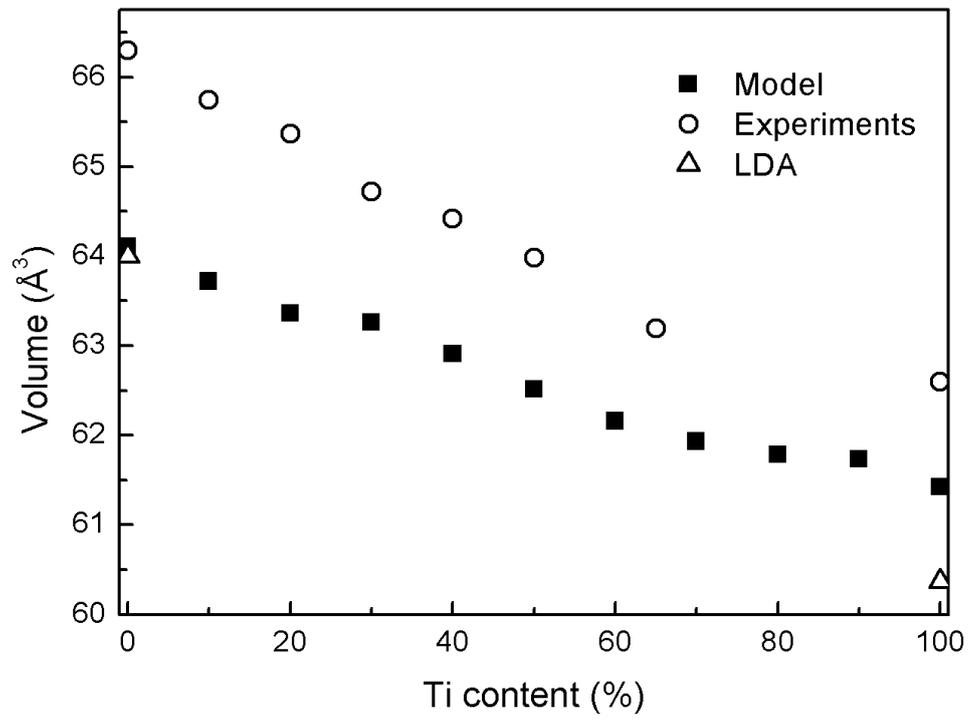

Figure 2

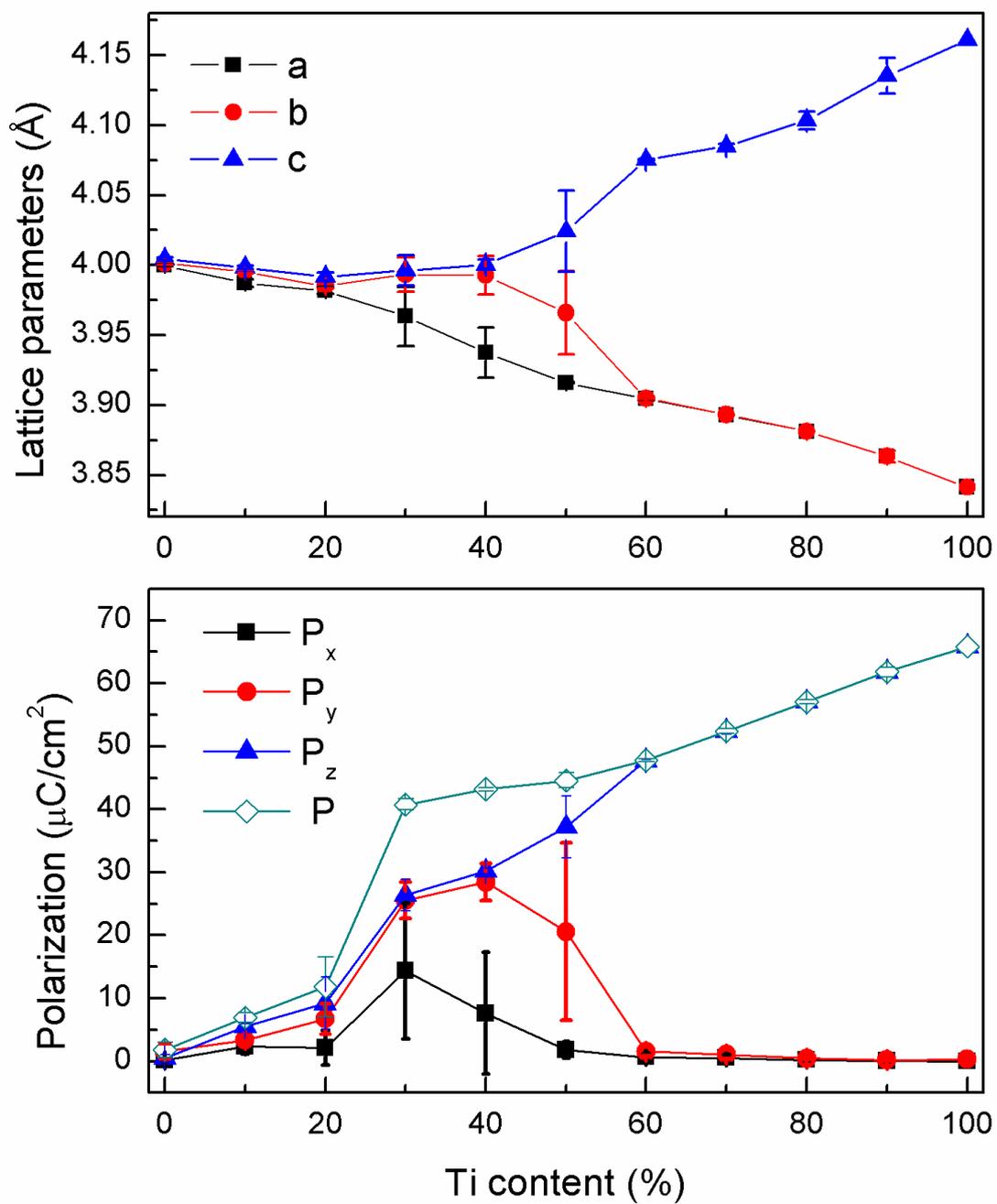

Figure 3

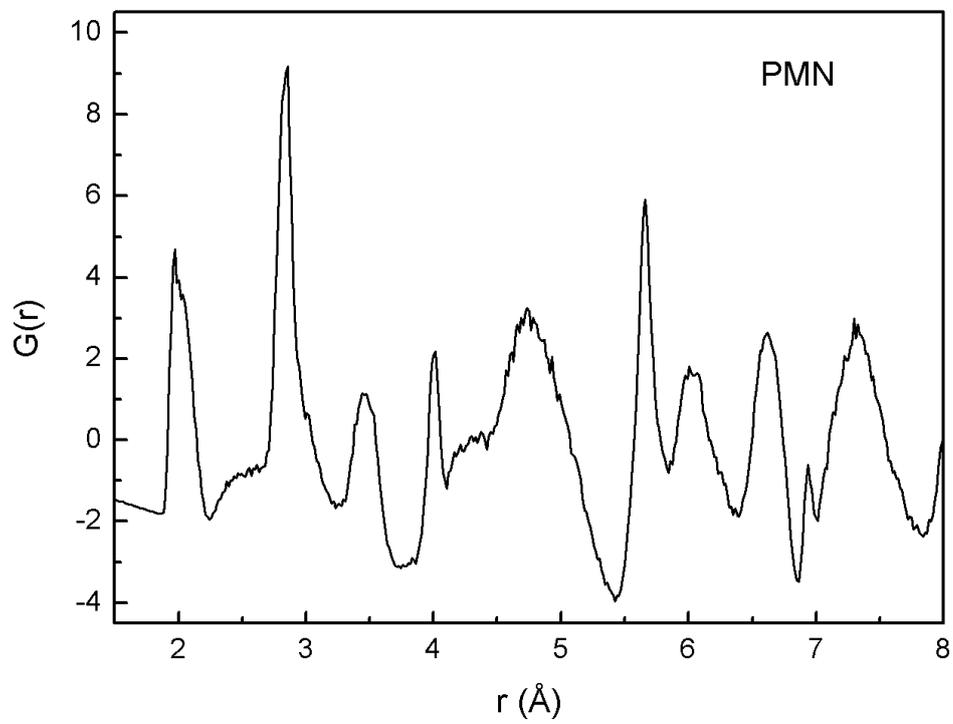

Figure 4

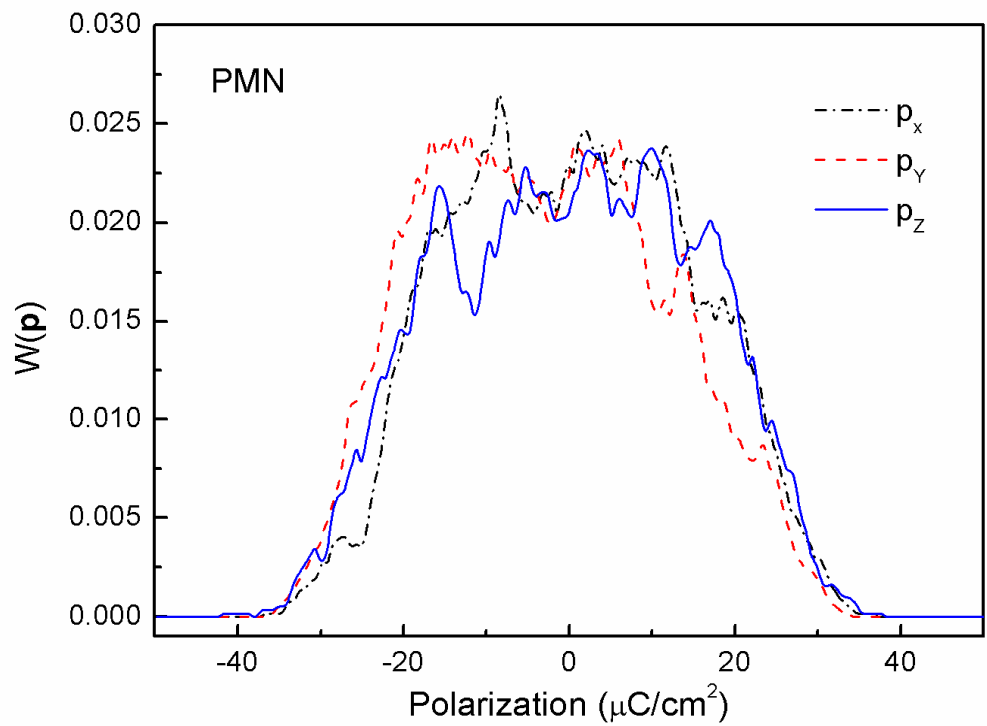

Figure 5

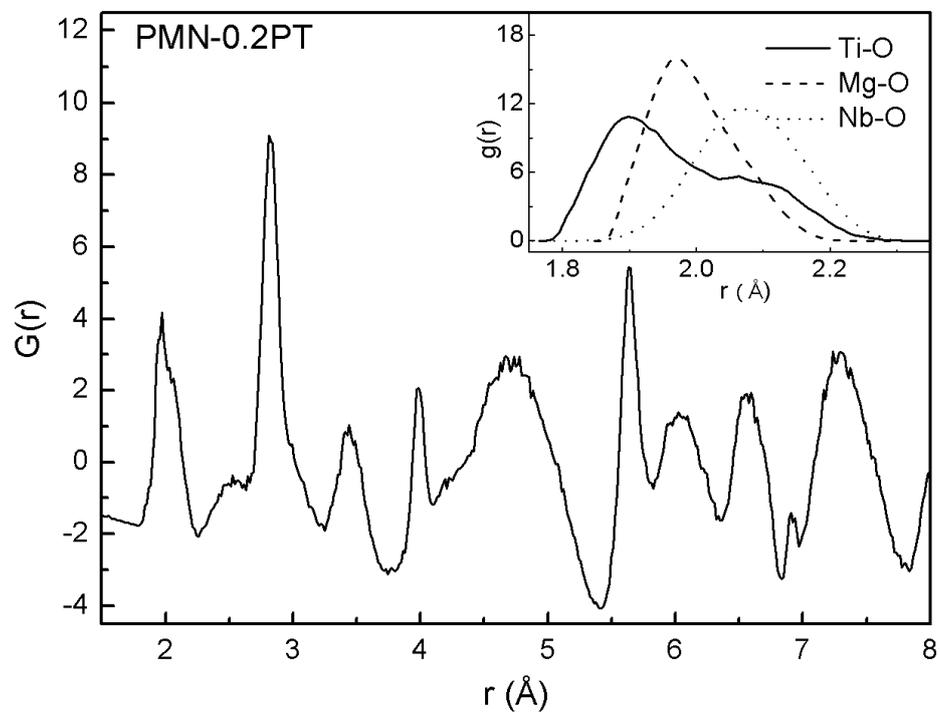

Figure 6

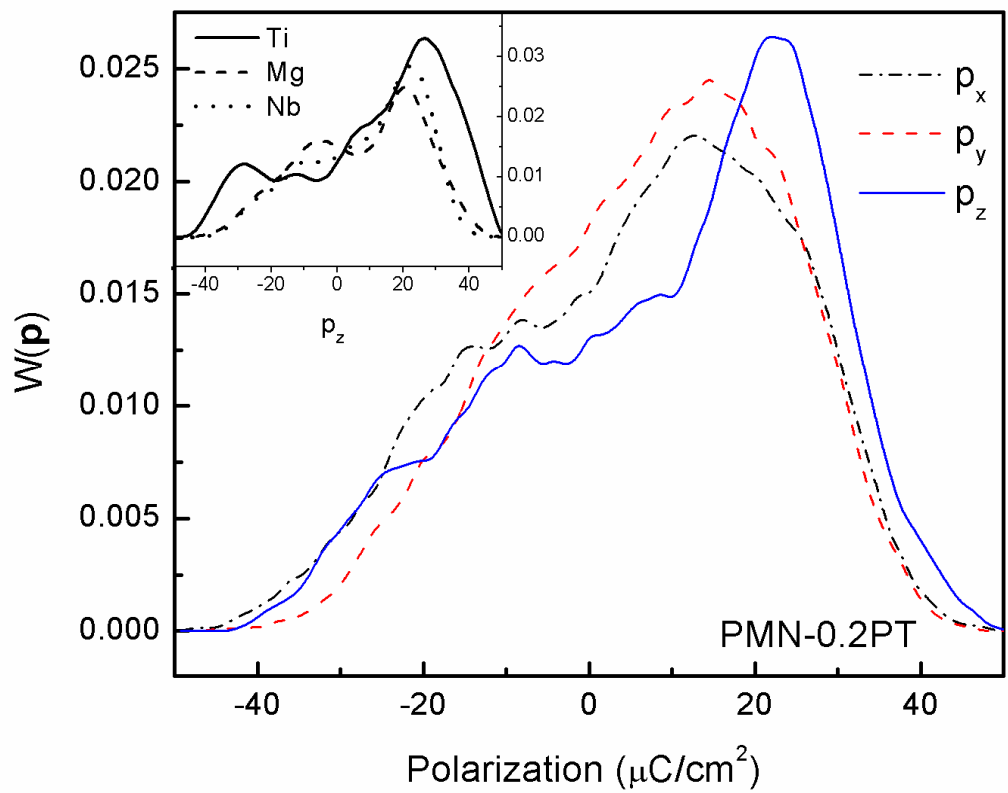

Figure 7

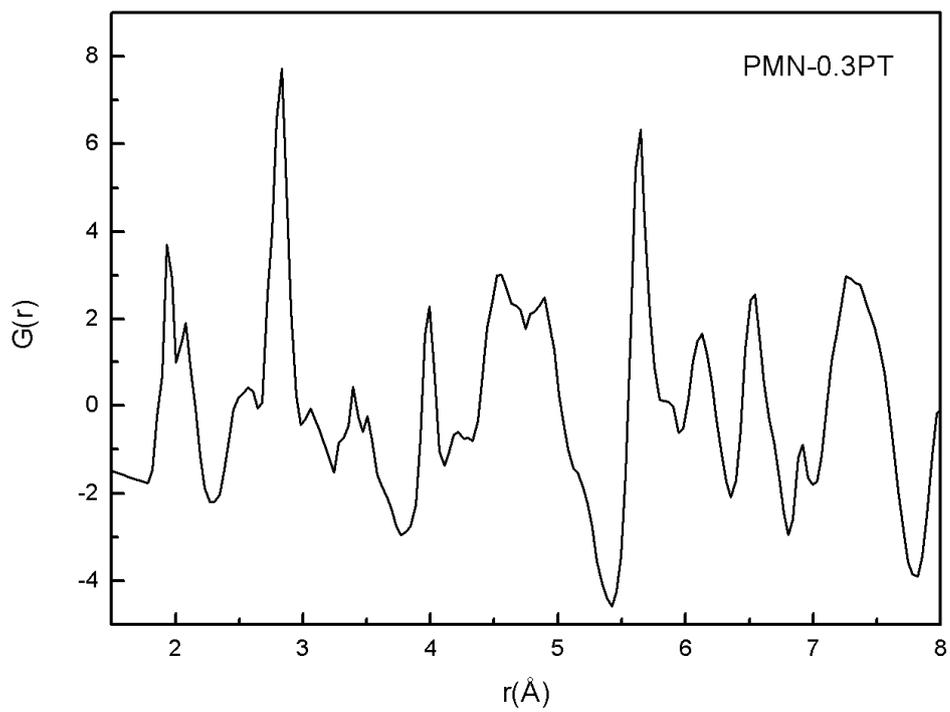

Figure 8

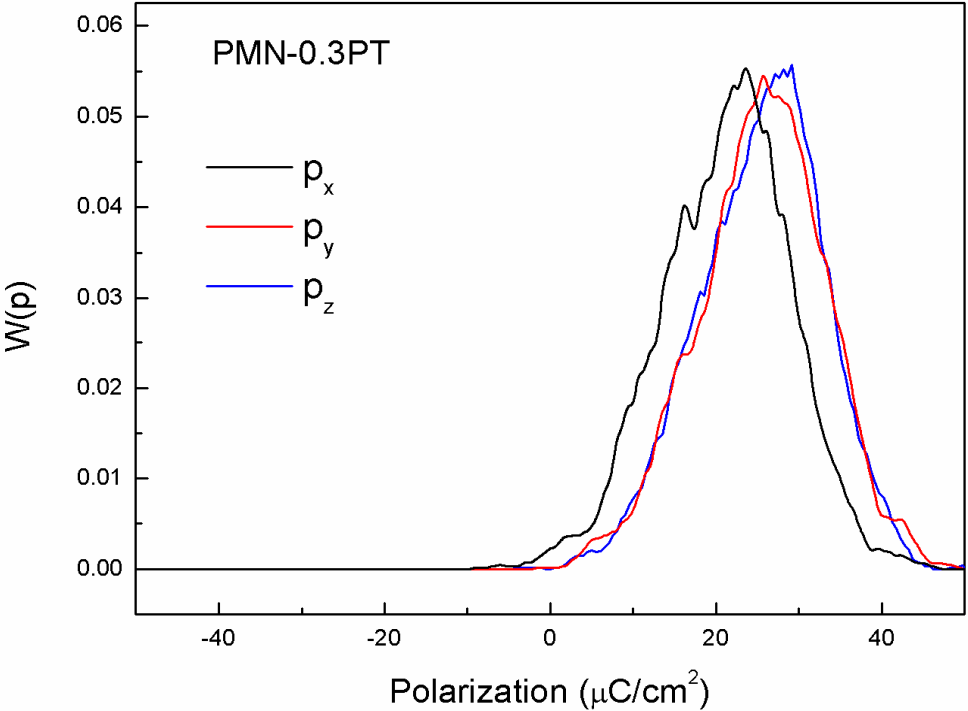

Figure 9

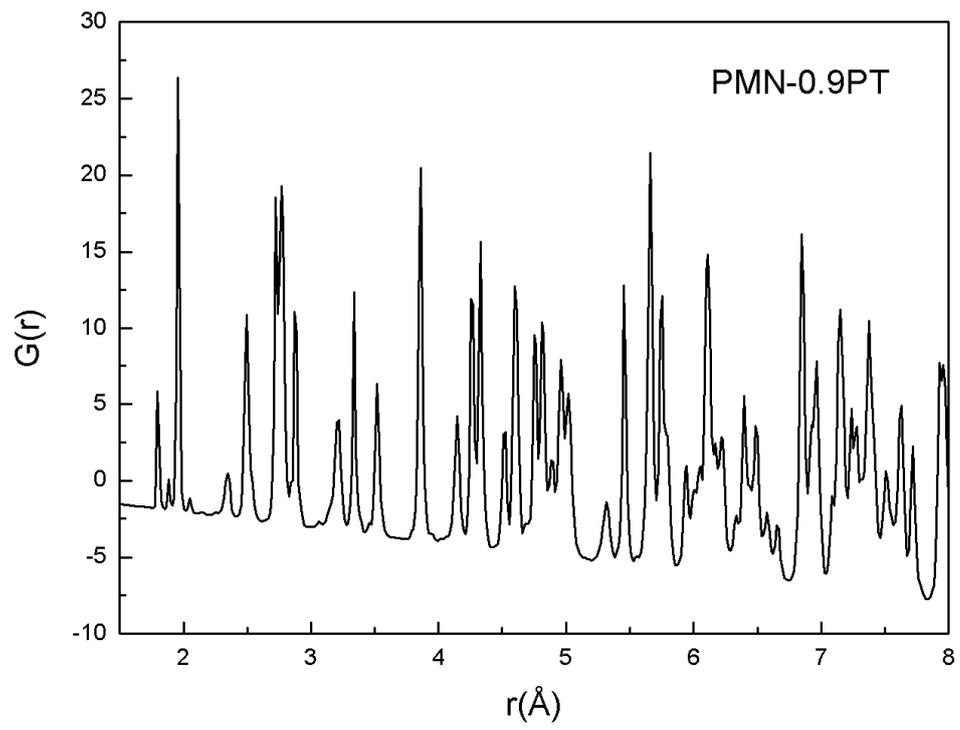

Figure 10

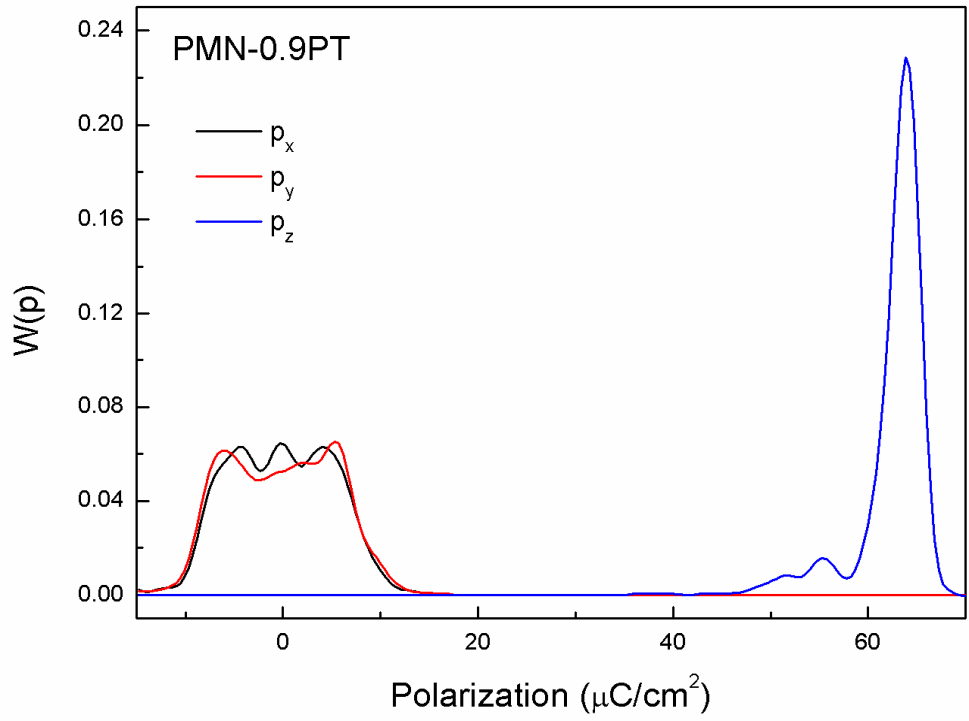